\documentclass[twocolumn,amssymb]{revtex4}
\newcommand{\singlespacing}{\let\CS=\@currsize\renewcommand{\baselinestretch}
{1.0}\tiny\CS}
\newcommand{\doublespacing}{\let\CS=\@currsize\renewcommand{\baselinestretch}
{1.5}\tiny\CS}

\begin{document}
\title{On the theory of topological computation in the lowest Landau level of QHE}
\author{Dipti Banerjee}
\thanks{Regular Associate of ICTP}
\email{email:deepbancu@homail.com,dbanerje@ictp.it}
\affiliation{Department of Physics,
Vidyasagar College for Women\\
$39$, Sankar Ghosh Lane,
Kolkata-$700006$, West Bengal\\
INDIA\\Tel:$+919830318243$}
\date{$5.5.09$}

\begin{abstract}
 We have studied the formation of Hall-qubit (LLL state)
in Quantum Hall effect due to the Aharonov-Bhom oscillation of
quasiparticles.The spin echo method plays the key role in the
topological entanglement of qubits. The proper ratio of fluxes for
maximally entangling qubits has also been pointed out. The
generation of higher Quantum Hall state may be possible with the
help of quantum teleportation.\\
\end{abstract}

\maketitle

Key words: spin echo, Aharanov-Bohm phase,.
PACS No:-$73.43$-f

Entanglement is one of the basic aspects of Quantum mechanics that
exhibits the peculiar correlations between two physically distant
parts of the total systems. The Geometric phase such as Berry
phase (BP) [1] and Aharanov-Bohm phase [2] play an important role in
Quantum mechanics. The effect of Berry phase on the entangled
quantum system is less known.The quantum gates that convey
topological transformation is known as topological gates.These
gates are advantageous for their immunity and resistive power for
local disturbances. They do not depend on the overall time
evolution nor on small deformations on the control parameter.This
indicates that a quantum mechanical state would carry its memory
during its spatial variation and the influence of Berry phase on
an entangled state could be linked up with the local observations
of spins.Few attempts have been made connecting the Berry phase
with entanglement of spin-$1/2$ particles resulting the outcome of
Geometric/Holonomic quantum computation [3].

The transport of information through quasi-particles exhibiting
long-range non-abelian Aharonov-Bohm interactions can yield
similar topological quantum computation [4]. Kitaev [5] pointed
out topological quantum computer as a device in which quantum
numbers carried by quasi-particles residing on 2DES have long
range Aharonov-Bohm interactions with one another.There exist a
strong quantum correlations between the quasi-particles
interacting by A-B effect extending out over large distances [6].

The interwinding of the quasi-particles trajectories in the course
of time evolution of the qubits realizes controlled-phase
transformations with nontrivial phase values.One of the remarkable
discoveries of recent decades is the infinite range A-B
interactions observed in Fractional Quantum Hall effect [7].
The electrons in the Quantum Hall systems are so highly frustrated
that the ground state is an extremely entangled state. The quantum
entanglement of particle having nontrivial Berry phase is
associated with the transport of a charge around a flux which is
equivalent to the Aharonov-Bohm phase in the analogy with the
gauge theory. A similar reflection is seen in FQHE when one
quasi-partcle/composite particle goes around another encircling an
area A. The total phase associated with this path is given by [8]
\begin{equation}
\Phi^{*}=-2\pi(BA/\phi_{0} - 2pN_{enc})
\end{equation}
where $N_{enc}$ is the number of composite fermion inside the
loop. The first term on the right hand side is usual A-B phase and
the second term is the contribution from the vortices bound to
composite fermions indicating that each enclosed composite fermion
effectively reduces the flux by $2p$ flux quanta. These particles
of a nontrivial condensed matter state obey fractional statistics
and Arovas et.al.[9] pointed out that the exchange of particles
over a half loop producing phase factor $e^{i\pi\theta}=
(-1)^{\theta}$.

In the entanglement of two spin-$1/2$ particles in presence of magnetic field through
the spin echo method BP plays an important role [10]. We have
studied in presence of Rabi oscillation the rotation of qubit
through BP [11]. A reflection of this
idea is seen to study the formation of Hall qubit
rotation recently in $\nu=1$ the ground state in the
light of quantum entanglement [12]. Here we aim at understanding
the rotation of Hall qubits having A-B interactions in the
background of topological computation.

\section{Birth of Hall-Qubit from Berry phase}

 One single qubit can be sufficiently constructed using the two well
known quantum gates - Hadamard gate(H) and Phase gate as follows
\begin{equation}
~~~|0>--[H]--{\bullet}^{2\theta}--[H]--{\bullet}^{\pi/2+\phi}\rightarrow \nonumber\\
\cos\theta|0>+\sin\theta
e^{i\phi}|1>
\end{equation}
With the use of quantum gates and elementary qubits $|0>$ and
$|1>$ the spinors for up or down polarization can be written as
\begin{equation}
\left|\uparrow\right\rangle
=\left(\sin\frac{\theta}{2}e^{i\phi/2}\left|0\right\rangle +
\cos\frac{\theta}{2}\left|1\right\rangle\right)
\end{equation}and
\begin{equation}
\left|\downarrow\right\rangle =
(-\cos\frac{\theta}{2}\left|0\right\rangle +
\sin\frac{\theta}{2}e^{-i\phi/2}\left|1\right\rangle)
\end{equation}

This above qubit representing a spinor acquires the geometric
phase [11] over a closed path.
\begin{eqnarray}
\gamma_{\uparrow}&=& i\oint\left\langle\uparrow,t\right|\nabla\left|\uparrow,t\right\rangle.d\lambda\\
 &=& i\oint A_{\uparrow}(\lambda)d\lambda\\
 &=& \oint L^\uparrow_{eff} dt\\
 &=& -\pi(1 - \cos\theta)
\end{eqnarray}
which is a solid angle subtended about the quantization axis. For
the conjugate state the Berry phase over the closed path becomes
\begin{equation}
\gamma_{\downarrow} = \pi(1-\cos\theta)
\end{equation}
Here the angle $\theta$ measures the deviation of the magnetic
flux line from the Z-axis. The fermionic or the antifermionic
nature of the two spinors (up/down) can be identified by the
maximum value of topological phase $\gamma_{\uparrow/
\downarrow}=\pm \pi$ at an angle $\theta=\pi/2$. For $\theta=0$ we
get the minimum value of $\gamma_{\uparrow}=0$ and at $\theta=\pi$
no extra effect of phase is realized.

In the spin echo method, this Berry phase can be fruitfully [17]
isolated in the construction of two qubit through rotation of one
qubit (spin 1/2) in the vicinity of another.Incorporating the
spin-echo for half period we find the antisymmetric Bell's state
after one cycle $(t=\tau)$,
\begin{equation}
\left|\Psi_- (t=\tau)\right\rangle =
\frac{1}{\sqrt{2}}(e^{i\gamma_\uparrow}\left|\uparrow\right\rangle_1\otimes
\left|\downarrow\right\rangle_2 -
e^{-i\gamma_\uparrow}\left|\downarrow\right\rangle_1
\otimes\left|\uparrow\right\rangle_2)
\end{equation}
and symmetric state becomes
\begin{equation}
\left|\Psi_+ (t=\tau)\right\rangle =
\frac{1}{\sqrt{2}}(e^{-i\gamma_\uparrow}\left|\uparrow\right\rangle_1\otimes
\left|\downarrow\right\rangle_2 +
e^{i\gamma_\uparrow}\left|\downarrow\right\rangle_1
\otimes\left|\uparrow\right\rangle_2)
\end{equation}
 where $\gamma_\downarrow = -\gamma_\uparrow =- \gamma$.
 It may be noted that for $|\mu_\uparrow|=\frac{1}{2}(1-cos\theta)$.

This phase plays the key role through the $\mu$ factor, in the
measurement of entanglement by the concurrence 'C' of an entangled
state. For a two qubit state
$|\psi>=\beta|\uparrow\downarrow>+\gamma|\downarrow\uparrow>$ the
concurrence is given by
\begin{equation}
 C=2\beta\gamma
\end{equation}
 When $C=1$ the entanglement is maximum and
disentanglement for $C=0$.
 The value of $|\mu_\uparrow|=0$ implies disentanglement for $\theta=0$.
 For $\theta=\pi$ there is maximum deviation of flux line
 yielding $|\mu_\uparrow|=1$ as a signature of maximum entanglement.

 Splitting up these above two eqs.(11) and (12) into the
 symmetric and antisymmetric states and rearranging we have
\begin{eqnarray}
\left|\Psi_+\right\rangle_\tau &=& \cos\gamma \left|\Psi_+\right\rangle_0 - i \sin\gamma \left|\Psi_-\right\rangle_0 \\
\left|\Psi_-\right\rangle_\tau &=& i \sin\gamma
\left|\Psi_+\right\rangle_0 + \cos\gamma
\left|\Psi_-\right\rangle_0
\end{eqnarray}
the doublet acquiring the matrix Berry phase $\Sigma$ as rotated
from $t=0$ to $t=\tau$.
\begin{equation}
{\left|\Psi_+\right\rangle \choose \left|\Psi_-\right\rangle}_\tau
= \Sigma {\left|\Psi_+\right\rangle \choose
\left|\Psi_-\right\rangle}_0
\end{equation}

\begin{equation}
\Sigma = \pmatrix{{\cos\gamma~~~-i\sin\gamma}\cr
{i\sin\gamma~~~~\cos\gamma}} = \cos 2\gamma
\end{equation}
This non-abelian matrix Berry phase $\Sigma$ is developed from the
abelian Berry phase $\gamma$. For $\gamma=0$ there is symmetric
rotation of states, but for $\gamma=\pi$ the return is
antisymmetric and the respective values of $\Sigma$=I and -I
(where I=identity matrix).

There is a deep analogy between FQHE and superfluidity [16]. The
ground state of anti-ferromagnetic Heisenberg model on a lattice
introduce frustration giving rise to the resonating valence
bond(RVB) states corresponding spin singlets where two
nearest-neighbor bonds are allowed to resonate among themselves.
The RVB state is a coherent superposition of spin singlet pairs
and can be written as
\begin{equation}
|RVB>= \sum(i_1 j_1,.. i_n j_n)\prod(i_k j_k)
\end{equation}
in which $(i,j)=\frac{1}{\sqrt{2}}(i\uparrow j\downarrow -
i\downarrow j\uparrow)$ is a spin singlet pair (valance bond)
between sites i and j. This RVB state support fractionalized
excitation of spin $1/2$ spinon [4,5].The topological order is
closely related to the coherent motion of fractionalised spin
excitation in RVB background. It is suggested that RVB states [6]
is a basis of fault tolerant topological quantum computation.
Since these spin singlet states forming a RVB gas is equivalent to
fractional quantum Hall fluid, its description through quantum
computation will be of ample interest.

The Quantum Hall effect can be considered on a 3D anisotropic
space having the N-particle wave-function of parent Hall states
\begin{equation} {\Psi_{N_\uparrow}}^{(m)}= \prod(u_i v_j - u_j
v_i)^m
\end{equation}
represent the one qubit in the language of quantum information.
 Following the Jain's formalism [15],${\Psi_\nu}^m$ the
hierarchical FQHE incompressible state for Landau filling factor $
\nu=\frac{p}{q}=\frac{n}{n(m-1) \pm 1}=\frac{2mn \pm 1}{n}$
becomes
\begin{equation}
|{\Psi_\nu (z)}^m > = |\Phi_n (z)>|\Phi_1 (z)>^{m-1}=|\Phi_1
(z)>^{2m+1/n}
\end{equation}
where $m$=odd for making the state ${\Psi_\nu (z)}^m$
antisymmetric and $n$=integer, specifying Lanadu level in QHE. The
state $\Phi_1(z)$ is the QHE state at the lowest landau level
$n=1$ and filling factor $\nu=1$. States of the above form are
grouped into a family depending on the values of $m$. Any FQHE
state can be expressed in terms of the IQHE ground state.

Recently we have identified [9] this ground state $\Phi_1(z)$ as
the Hall qubit, the basic building block for constructing any
other IQHE/FQHE state formed when two nearest neighbor bonds are
allowed to resonate among themselves. It is an extremely entangled
state visualized by the formation of singlet state between a pair
of $(i,j)^{th}$ spinors.
\begin{eqnarray}
\Phi_1 (z) &=& (u_i v_j - u_j v_i)\nonumber\\
&=&(u_i~~~~v_i)\pmatrix {{0~~~~~1}\cr{-1~~~~~0}}{u_j \choose
v_j}\\
&=&\langle\uparrow_i| \pmatrix
{{0~~~~~1}\cr{-1~~~~~0}}|\uparrow_j\rangle
\end{eqnarray}
A singlet state is a two qubit developed as one qubit
$\left|\uparrow\right\rangle = {u \choose v}=
{\sin\frac{\theta}{2}e^{i\phi} \choose \cos\frac{\theta}{2}}$
rotates in the vicinity of another through the spin echo method
where Berry's topological phase dominates in acquiring the Hall
qubit with the description of a two component up spinor [14]

In this present work we have focused on Quantum Hall effect where
the nature of the state will be only antisymmetric. Hence the
eq.(17) reduces to
$${\left|\Psi_-\right\rangle}_\tau=\Sigma{\left|\Psi_-\right\rangle}_0$$
 The Hall qubit $\Phi_1 (z)$ has resemblance with $\left|\Psi_-\right\rangle$.
 In the lowest Landau level $\nu=m=1$ would develop similar non-abelian Berry
phase $\Sigma$. This is visualizing the spin conflict during
parallel transport leading to matrix Berry phase [9]. Over a
closed period $t=\tau$ the QHE state $\Phi_1(z)$ at $\nu=1$
filling factor will acquire the matrix Berry phase.
\begin{equation}
\left|\Phi_1 (z)\right\rangle_\tau =
e^{i{\gamma^H}_{\uparrow}}\left|\Phi_1 (z)\right\rangle_0
\end{equation}
Here $e^{i\gamma^H}$ is the non-abelian matrix Berry phase
\begin{equation}
{\gamma^H}_{\uparrow}=\pmatrix{{\gamma_i~~~~\gamma_{ij}}\cr{\gamma_{ji}~~~~\gamma_j}}
\end{equation}
 where $\gamma_i$ and $\gamma_j$ are the BPs for the ith
and jth spinor as seen in eq.(16) and the off-diagonal BP
$\gamma_{ij}$ arises due to local frustration in the spin system.

All the above explanation is restricted for lowest Landau level
$\nu=1$, but concentrating on the other parent state $\nu=1/m$
where $m=$odd integers, the Berry phase of a qubit is $\gamma=i m
\pi$ [14] that is associated with the two qubit Hall state
\begin{equation}
\left|\Phi_1 (z)\right\rangle
=\frac{1}{\sqrt{2}}(\left|\uparrow\right\rangle_1~~\left|\downarrow\right\rangle_1)\pmatrix
{{0~~~-e^{-im\pi}}\cr{e^{im\pi}~~~0}}{\left|\uparrow\right\rangle_2
\choose\left|\downarrow\right\rangle_2}
\end{equation}
through the process of quantum entanglement between two one qubit
where the reflection of spin echo is visualized.

\section{The Qubit formation in QHE through Aharanov-Bohm phase}

In the composite fermion theory of Quantum Hall effect the qubits
are equivalent to the fluxes attached with the charged particles.
When an electron  is attached with a magnetic flux, its statistics
changes and it is transformed into a boson.These bosons condense
to form cluster which is coupled with the residual fermion or
boson (composed of two fermions). Indeed the residual boson or
fermion will undergo a statistical interaction tied to a geometric
Berry phase effect that winds the phase of the particles as it
encircles the vortices.Indeed as two vortices cannot be brought
very close to each other, there will be a hard core repulsion in
the system which accounts for the incompressibility of the Quantum
Hall fluid.

These non-commuting fluxes have their own interesting
Aharonov-Bohm interactions.As the quasi-particles encircles
another in their way of topological transport, the Aharanov-Bhom
type statistical phase is developed. Following a generalization of
Pauli exclusion principle, Haldane[12] pointed out that the
quasi-particles carrying flux $\phi_{\alpha}$ and charge
$q_{\beta}$ orbiting around another object carrying flux
$\phi_{\beta}$ and charge $q_{\beta}$ has the relative statistical
phase $\theta_{\alpha\beta}$
\begin{equation}
exp(i\theta_{\alpha\beta}) = exp\pm i\pi
(g_{\alpha\beta}+g_{\beta\alpha})
\end{equation}~~~~~~~~~~~~~~~~~where
$g_{\alpha\beta}=-q_{\alpha}\phi_{\beta}$.
 With this view we have recently shown that when two
 non-identical composite fermions residing in two consecutive
 Landau levels in FQHE encircle each
other, the relative Aharonov-Bhom (AB) type phase is developed. As
the quasi-particles advance towards the edge of FQHE in a similar
circular way, the developed current [13] should have a connection
with this AB type phase through Berry's topological phase.

 This A-B interactions are the
key source of forming two qubit Hall states identified as Hall
qubit. Hence movement of Hall qubits would develop the A-B phase.
In the physics of spin echo instead of Berry phase the
incorporation of Aharonov-Bhom phase would be more appropriate as
the rotation of qubits are equivalent to the rotation of fluxes
around charges. If $e^{i\phi_s}$ be the Aharanov-Bohm phase
between the two qubits, for half period we find the antisymmetric
Bell's state after one cycle $(t=\tau)$,
\begin{equation}
\left|\Psi_- (t=\tau)\right\rangle =
\frac{1}{\sqrt{2}}(e^{i\phi_s}\left|\uparrow\right\rangle_1\otimes
\left|\downarrow\right\rangle_2 -
e^{-i\phi_s}\left|\downarrow\right\rangle_1
\otimes\left|\uparrow\right\rangle_2)
\end{equation}
similar consideration for symmetric states
\begin{equation}
\left|\Psi_+ (t=\tau)\right\rangle =
\frac{1}{\sqrt{2}}(e^{-i\phi_s}\left|\uparrow\right\rangle_1\otimes
\left|\downarrow\right\rangle_2 +
e^{i\phi_s}\left|\downarrow\right\rangle_1
\otimes\left|\uparrow\right\rangle_2)
\end{equation}
Splitting up these states and rearranging the symmetric and
antisymmetric parts we have the doublet acquiring the matrix form
of Aharonov-Bhom phase $\Upsilon$ as rotated from $t=0$ to
$t=\tau$.
\begin{equation}
{\left|\Psi_+\right\rangle \choose \left|\Psi_-\right\rangle}_\tau
= \Upsilon {\left|\Psi_+\right\rangle \choose
\left|\Psi_-\right\rangle}_0
\end{equation}
where
\begin{equation}
\Upsilon = \pmatrix{{\cos\phi_s~~~-i\sin\phi_s}\cr
{i\sin\phi_s~~~~\cos\phi_s}} = \cos 2\phi_s
\end{equation}
This topological matrix phase $\Upsilon$ is developed from the
Aharonov-Bohm phase $\phi_s$ as one qubit rotates around another.
The qubits in QHE are quantized spinor having flux attached with
charge. Their entanglement is equivalent to spin type echo where
the topological phase dominates due to Aharonov-Bohm oscillation
between them.
 This compel to change the
Berry phase of the singlet state as in eq.(24) by the relative A-B
type phase $\phi_s$.
\begin{equation}
\left|\Phi_1 (z)\right\rangle_\tau = e^{i\phi_s}\left|\Phi_1
(z)\right\rangle_0
\end{equation}
This Hall qubit can be visualized in terms of entanglement of two
 oscillating qubits .
\begin{equation}
\left|\Phi_1 (z)\right\rangle
=\frac{1}{\sqrt{2}}(\left|\uparrow\right\rangle_1~~\left|\downarrow\right\rangle_1)\pmatrix
{{0~~~-e^{-i\phi_s}}\cr{e^{i\phi_s}~~~0}}{\left|\uparrow\right\rangle_2
\choose\left|\downarrow\right\rangle_2}
\end{equation}
To form the singlet state between the qubits under A-B
interactions in the spin echo method, the essential condition for
 antisymmetric QHE states are visualized by $\pm e^{i\phi_s}=\pm e^{i\pi}=\pm
1$.

In the formation of Hall qubit through the entangling of qubits in
the different Landau level the A-B phase plays the key role in the
spin type echo method. If the rotating qubits are in the same
Landau level, the A-B phase changes to statistical phase.
Considering the interaction [18] between two identical qubits in
the same $n^{th}$ Landau level for the composite particles filling
factor $\nu=\frac{n}{2\mu_{eff}}$ the statistical phase becomes
\begin{equation}
\phi_s=exp \pm i\frac{n\pi}{2}
\end{equation}
where for $n=2,4,6...$ the change of statistics will be fermionic
visualized by the phase $\phi_s=exp \pm i\pi$.On the other hand
for $n=1,3,5..$ the statistics will be bosonic $\phi_s=exp \pm
i\pi/2$. It may be noted that for fractional filling factors the
final statistics will be fermionic through proper combinations.
With this view, the entanglement of two one qubits in the same
Landau Level, the Hall qubit $\left|\Phi_1 (z)\right\rangle$ will
be formed when the exchange phase is $\pm i\pi$
\begin{equation}
\left|\Phi_1 (z)\right\rangle
=\frac{1}{\sqrt{2}}(\left|\uparrow\right\rangle_1~~\left|\downarrow\right\rangle_1)\pmatrix
{{0~~~-e^{-i\pi}}\cr{e^{i\pi }~~~0}}{\left|\uparrow\right\rangle_2
\choose\left|\downarrow\right\rangle_2}
\end{equation}

Whenever the interaction takes place between dissimilar qubits in
different Landau level the rotation of one against another
develops the Aharanov-Bhom type phase that does not express their
statistics. We assumed the transfer of the composite particle [13]
from the inner edge in the $n^{th}$ Landau level having filling
factor $\nu_n$ picking up even integral $(2m)$ of flux $\nu_1$
through the bulk of QH system and forming a new composite particle
in the $(n+1)^{th}$ Landau level in the outer edge. The filling
factor of the effective particle becomes
$\nu_{eff}=\frac{n+1}{\mu_{eff}}$. The monopole strength
$\mu_{eff}$ of the state ${\Phi_{1}}^{2m}\Phi_n$ can be considered
as
\begin{equation}
\mu_{eff}=2m\mu_1+\mu_n.
\end{equation}
  Encircling of the composite particle
 in the inner edge having flux $\mu_n$ with charge $q_n$
 around the composite particle in the outer edge having corresponding flux
 $\mu_{eff}$ would develop a relative AB type phase
\begin{equation}
\phi_s=exp \pm{\frac{i\pi}{2}}(q_n\mu_{eff}+q_{eff}\mu_n)
\end{equation}
In more simplified way it becomes
\begin{equation}
\phi_s \cong exp\pm{\frac{i\pi}{2}}[(n+\frac{1}{2})- m
\frac{\mu_1}{\mu_n}]
\end{equation}
 Since the concurrence $C=1$ indicate the maximum entanglement and for
disentanglement the value of minimum concurrence is $C=0$, we can
establish a relationship between the fluxes of the entangling
qubits on the Hall surface.The maximum entanglement between the
two quasi-particle results a relation between the respective two
fluxes $\mu_1$ and $\mu_n$ in terms of parent filling factor $m$
and Landau level $n$.
\begin{equation}
 \exp\pm{\frac{i\pi}{2}}[(n+\frac{1}{2})- m
\frac{\mu_1}{\mu_n}]= \exp\pm i\pi
\end{equation}
This gives a ratio between the entangling fluxes $\mu_n$ and
$\mu_1$ in order to form the singlet pairs through AB oscillations
in Quantum Hall effect.
\begin{equation}
\mu_n=\frac{2m}{2n-3} \mu_1
\end{equation}
 It may be noted that for maximum entanglement $\mu_1=1$ results
$$\mu_n=\frac{2m}{2n-3}$$ and for minimum entanglement both
$\mu_1$ and $\mu_n$ becomes zero.

The physics behind the formation of higher Hall states take place
through the entanglement of Hall qubits $|\Phi(z)_1>$ in the
lowest landau level. Here the A-B phase or statistical phase plays
the key role in the process of spin echo with the essential
condition $\pm e^{i\phi_s}=\pm e^{i\pi}=\pm 1$. The outcome of
entanglement of two Hall qubits is
\begin{equation}
|\alpha(z)>=<\Phi_1 (z)|\pmatrix {{0~~~-e^{-i\pi}}\cr{e^{i\pi
}~~~0}}|\Phi_1 (z)>= \pmatrix {{0~~~\Phi_1 (z)}\cr{-\Phi_1
(z)~~~0}}
\end{equation}
where $\Phi_1 (z)=(u_i v_j - v_i u_j)$. On the similar manner we
realize that the entanglement of two $|\alpha(z)>$ gives rise the
state formed by the square of Hall qubits.
\begin{equation}
|\gamma(z)>=<\alpha (z)|\pmatrix{{0~~~1}\cr{-1~~~0}}|\alpha
(z)>=\pmatrix {{0~~~{\Phi_1}^2 (z)}\cr{-{\Phi_1}^2 (z)~~~0}}
\end{equation}
In order to maintain the antisymmetric nature of the Hall state
the power of the Hall qubit should be odd. This is possible only
when two asymmetric Hall qubits ( one even power with another odd
power) entangle under topological interactions
\begin{equation}
<\gamma(z)|\pmatrix{{0~~~1}\cr{-1~~~0}}|\alpha (z)>=\pmatrix
{{0~~~{\Phi_1}^3 (z)}\cr{-{\Phi_1}^3 (z)~~~0}}
\end{equation}
forming a Hall state in the parent Landau filling factor
$\nu=m=odd$.

 We like to conclude mentioning that the hierarchical FQHE
states are formed through the process of quantum teleportation. If
there are three entities defined by 1,2, and 3, then
transportation of 1 to 3 through 2 will be [19]
\begin{eqnarray}
|\Psi>_{123}&=& |\Phi_1>|\Psi_{23}>\\ &=&
\frac{1}{2}(1+\sigma^1.\sigma^3)|\Psi_{12}>|\Phi_3>
\end{eqnarray}
Similar reflection of quantum teleportation in FQHE motivate us to
write
\begin{equation}
{\Psi_\nu}^m={\Phi_{1}}^{2m}\Phi_n={\Phi_{1}}^{2m}
{\Phi_1}^{\frac{1}{n}}=\frac{1}{2}(1+\sigma^1.\sigma^3)|{\Phi_1}^{\frac{1}{n}}
{\Phi_1}^{2m}
\end{equation}
a hierarchical FQHE state whose extensive study through the
entanglement of Hall qubits maintaining the antisymmetric nature
of the exchange phase is to be done in future.
\section{Discussion}
 In this paper we have studied the Physics behind the
Hall qubit formed by the entanglement of two qubits where one is
rotating in the field of the other with Aharonov-Bohm (AB)
phase.The image of spin echo between the entangling composite
fermion/qubit has been reflected in the field of Quantum Hall
effect.Topological quantum computation has been executed
considering the Hall qubit at $\nu=1$ as a building block
for the formation of
other higher IQHE/FQHE states at different filling factors.
 Through the
condition of concurrence for maximum entanglement, a proper ratio
between the fluxes of the entangling qubits has been evaluated.
At the end, we have mentioned that the states in hierarchies of
FQHE can be studied in the light of quantum teleportation whose
extensive study will be of ample interest in future.\\
{\bf Acknowledgements:}\\
This work has been partly supported by The Abdus Salam
International Center for Theoretical Physics, Trieste, Italy.

\end{document}